\documentstyle[prl,aps,epsfig,twocolumn]{revtex}

\begin{document}

\draft

\twocolumn[\hsize\textwidth\columnwidth\hsize
\csname@twocolumnfalse\endcsname
\title{Non-Abelian Geometric Quantum Memory with Atomic Ensemble}
\author{Y. Li $^{1}$, P. Zhang $^{1}$, P. Zanardi $^{2,3}$, and C. P. Sun $^{1,a,b}$}
\address{$^{1}$ Institute of Theoretical Physics, The Chinese Academy of Science, Beijing, 100080,
China\\
{$^2$ Department of Mechanical Engineering, Massachusetts Institute
of Technology, Cambridge, MA 02139}
\\
$^{3}$ Institute for Scientific Interchange (ISI), Viale Settimio
Severo 65, 10133 Torino, Italy}
 \maketitle

\begin{abstract}
We study a quantum information storage scheme based on an atomic
ensemble with near (also exact) three-photon resonance
electromagnetically induced transparency (EIT). Each 4-level-atom
is coupled to two classical control fields and a quantum probe
field. Quantum information is adiabatically stored in the
associated dark polariton manifold. An intrinsic non-trivial
topological structure is discovered in our quantum memory
implemented through the symmetric collective atomic excitations
with a hidden $SU(3)$ dynamical symmetry. By adiabatically changing
the Rabi frequencies of two classical control fields, the quantum
state can be retrieved up to a non-abelian holonomy and thus
decoded from the final state in a purely geometric way.
\end{abstract}

 ]

Quantum information storage is a physical process to encode the
state of a quantum system into the state of another system referred
to as a quantum memory \cite{q-infor}. Compared to the original
quantum system the quantum memory should possess a large
decoherence time for effective storing of quantum information.
Moreover, the original state of the quantum system should be
retrievable from the encoding quantum memory state. By means of
quantum memory one can transport quantum information from place to
place within the decoherence time. Recently ensemble of
$\Lambda$-type atoms has been proposed
\cite{Lukin00-ent,Fl00-pol,Fl00-OptCom} as a candidate for
practical quantum memory. The idea is to store and transfer quantum
information contained in photonic states by the collective atomic
excitations. This approach is based on the phenomenon of
electromagnetically induced transparency (EIT) \cite{EIT}. Some
experiments \cite{lui,group} have already demonstrated the central
principle of this technique, namely, the reduction of the group
velocity of light.

Most recently a system with quasi-spin wave collective excitations
of many $\Lambda$-type atoms fixed in "atomic crystal" has been
considered as a candidate for a robust quantum memory
\cite{Sun-prl}. A hidden dynamical symmetry of such a system is
discovered and it is observed that in certain cases
\cite{Sun-quant-ph} the quantum state can be retrieved up to a
non-abelian Berry phase, i.e., a non-abelian holonomy
\cite{BPF,ZW,Zana,Un,duan-science,Ekert}. This observation extends
the concept of quantum information storage. Quantum information
storage of photonic states with this topological character can be
implemented in an atomic ensemble with off-resonance EIT. In such a
case the stored state can be decoded in a purely geometric way.
However, this non-abelian holonomy is in some sense trivial due to
the fact that the quantum storage space splits into an orthogonal
sum of invariant one dimensional subspaces.

In this letter, we shall describe a quantum information storage
protocol based on a truly non-abelian holonomy. To this aim we will
consider an ensemble of $N$ 4-level-atoms \cite{Un,duan-science},
where two meta-stable states are coupled to the excited state by
two classical control fields respectively while the ground state is
coupled to the excited state by a quantum probe field. In the large
$N$ limit with low excitation, a three-exciton system is formed by
the symmetric collective excitations from the ground states up to
the excited state plus the two virtual excitations from the two
meta-stable states to the excited state. It is easy to prove that
these three collective excitations indeed behave as three bosons in
the large $N$ limit with low excitation. Intertwining between the
excited state and two meta-stable ones, the collective operators
generate an $SU(3)$ algebra. Based on the spectrum generating
algebra theory \cite{algebra} associated with this $SU(3)$, we
construct the degenerate eigen-states of the three-mode
exciton-photon system. In particular the collective manifold of
dark states can be shown to split into dynamically invariant
higher-dimensional subspaces. Using these degenerate eigen-states
as a quantum memory, quantum information storage of photonic states
can be implemented up to a non-abelian holonomy.
\begin{figure}[h]
\hspace{30pt}\includegraphics[width=4cm,height=4cm]{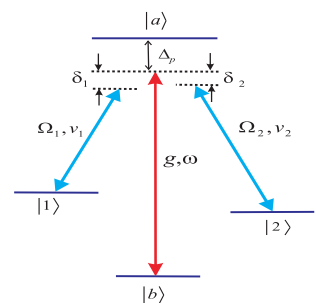}
\caption{Four-level atom interacting with a quantum probe field
(with coupling constant $g$, frequency $\protect\omega$, and the
detuning $ \Delta_{p}$) and two classic control fields (with
frequency $\protect\nu_{k}$ , coupling Rabi frequency $\Omega_{k}$,
and the detuning $\Delta_{k}=\protect \omega
_{ak}-\protect\nu_{k}$, $k=1,2$). When $\protect\delta_{k}$
$(=\Delta _{k}-\Delta _{p})$ are very tiny, the system satisfies
the near 3-photon resonance EIT condition.}
\end{figure}
Our system consists of $N$ identical 4-level atoms
\cite{Un,duan-science}, where all the atoms are coupled to two
single-mode classical control fields and a quantum probe field as
shown in Fig. 1. The atomic levels are labelled as the ground state
$|b\rangle$, the excited state $|a\rangle$, and the meta-stable
states $|k\rangle$ $(k=1,2)$. The atomic transition
$|a\rangle\rightarrow|b\rangle$, with energy level difference
$\omega_{ab}$=$\omega_{a}-\omega_{b}$, is coupled to the probe
field of frequency $\omega$ $(=\omega_{ab}-\Delta_{p})$ with the
coupling coefficient $g$; and the atomic transition
$|a\rangle\rightarrow|k\rangle$ $(k=1,2)$, with energy level
difference $\omega_{ak}$, is driven by the classical control field
of frequency $\nu_{k}$ $(=\omega_{ak}-\Delta_{k})$ with
Rabi-frequency $\Omega_{k}(t)$.

In the present work we consider the case of $\delta_{k}$
$(=\Delta_{k}-\Delta_{p})$ being very small, that is, those three
fields have almost the same detuning with respect to the upper
level $|a\rangle$. In view of the physical intuition, each
meta-stable state with its relevant control field would constitute
a near two-photon resonance EIT if another meta-stable state and
its relevant control field do not exist. With the case of
two-photon resonance EIT \cite{Lukin-RMP,Deng1,Deng2} (where the
control and probe fields have the same detuning) in mind, we would
refer to our case of $\Delta_{p}\simeq\Delta_{k}$ as a {\em near}
{\em "3-photon resonance" EIT}.

Under the rotating wave approximation the interaction Hamiltonian
can be written as (let $\hbar =1$) \cite{Sun-prl}
\begin{eqnarray}
H_{I} &=&\Delta _{p}S+g\sqrt{N}aA^{\dagger }+\Omega _{1}\exp [i\phi
_{1}(t)]T_{+}^{(1)}  \nonumber \\
&&+\Omega _{2}\exp [i\phi _{2}(t)]T_{+}^{(2)}+h.c.,
\end{eqnarray}
where
\begin{eqnarray}
S &=& \sum_{j=1}^{N}\sigma _{aa}^{(j)},\text{ }A=\frac{1}{\sqrt{N}}%
\sum_{j=1}^{N}\sigma _{ba}^{(j)},  \nonumber \\
T_{-}^{(k)} &=& \sum_{j=1}^{N}\sigma _{ka}^{(j)},\text{ }%
T_{+}^{(k)}=(T_{-}^{(k)})^{\dagger },\text{ }k=1,2,  \label{coll-ops}
\end{eqnarray}%
are symmetrized collective atomic operators. Here $\sigma _{\mu \nu
}^{(j)}=|\mu \rangle _{jj}\langle \nu |$ denotes the flip operator
of the $j$-th atom from state $|\nu \rangle _{j}$ to $|\mu \rangle
_{j}$ $(\mu ,\nu =a,b,1,2)$; $a^{\dagger }$ and $a$ the creation
and annihilation operators of quantum probe field respectively; and
$\phi _{k}(t)=$ $\delta _{k}t$. The coupling coefficients $g$ and
$\Omega _{1,2}$ are real and assumed to be identical for different
atoms in the ensemble. A similar effective Hamiltonian was given in
Ref. \cite{Sun-prl} for the case of an "atomic crystal", in terms
of quasi-spin-wave type collective atomic operators and a hidden
dynamical symmetry was there discovered. The symmetrized operators
(\ref{coll-ops}) are just a special instance of the quasi-spin-wave
operators discussed in \cite{Sun-prl}.

Let us first consider a similar dynamical symmetry in the low
excitation regime of the atomic ensemble where most of $N$ atoms
stay in the ground state $|b\rangle$ and $N\rightarrow\infty$. It
is obvious that $T_{-}^{(k)}$ and $T_{+}^{(k)}$ $(k=1,2)$ generate
two mutually commuting $SU(2)$ subalgebras of $SU(3)$
\cite{Sun-prb}. To form a closed algebra containing $SU(3)$ and
$\{A,A^{\dagger}\}$, we need to introduce two additional collective
operators
\begin{equation}
C_{k}=\frac{1}{\sqrt{N}}\sum_{j=1}^{N}\sigma_{bk}^{(j)},\text{ }k=1,2
\label{C_k}
\end{equation}
along with their hermitian conjugates. These operators have the
non-vanishing commutation relations $C_{k}=[A,T_{+}^{(k)}]$ and $
[C_{k},T_{-}^{(k)}]=A$ $(k=1,2)$. As a special case of quasi-spin
wave excitation with zero varying phases, the above three mode
symmetrized excitations defined by $A$ and $C_{1,2}$ behave as
three independent bosons. Indeed one can check that the operators
(\ref{C_k}), in the large $N$ limit with low excitation, satisfy
the bosonic commutation relations \cite{Sun-prb}. The commutation
relations between the $SU(3)$ algebra and the Heisenberg-Weyl
algebra $h$ generated by $A$, $A^{\dagger}$, $C_{k}$, and
$C_{k}^{\dagger}$ imply that the dynamical symmetry of evolution
governed by $H_{I}$ can be described by the semi-direct product
algebra $SU(3)\overline{\otimes}h$.

Based on this hidden dynamical symmetry of the interaction
Hamiltonian, we can introduce a dark-state polariton operator
\begin{equation}
D=a\cos\theta-C\sin\theta,
\end{equation}
where
\begin{equation}
C=C_{1}\exp[i\phi_{1}(t)]\cos\kappa+C_{2}\exp[i\phi_{2}(t)]\sin\kappa
\end{equation}
is a coherent mixing of two collective atomic excitations $C_{1}$
and $C_{2}$, and $\kappa=\arctan\frac{\Omega_{2}}{\Omega_{1}}$,
$\theta=\arctan \frac{g\sqrt{N}}{\Omega}$,
$\Omega=\sqrt{\Omega_{1}^{2}+\Omega_{2}^{2}}$. In terms of a new
operator by
\begin{equation}
T_{+}=T_{+}^{(1)}\exp[i\phi_{1}(t)]\cos\kappa+T_{+}^{(2)}\exp[i\phi
_{2}(t)]\sin\kappa,
\end{equation}
we can then rewrite the interaction Hamiltonian as
\begin{equation}
H_{I}=\Delta_{p} S+g\sqrt{N}aA^{\dagger}+\Omega T_{+}+h.c..
\end{equation}
Since $[C,T_{-}]=A$ and $[A,T_{+}]=C$, one can readily verify that
$[D,H_{I}]=0$. To generate the full eigen-space of $H_{I}$ with
zero eigenvalue, i.e., the dark-polariton manifold, we need to
consider another dark-state polariton operator complementary to
$D$:
\begin{equation}
E=C_{2}\exp[i\phi_{2}(t)]\cos\kappa-C_{1}\exp[i\phi_{1}(t)]\sin\kappa.
\label{E02}
\end{equation}
It is worthwhile to point out that $E$ satisfies the bosonic
commutation relation as well and it is independent of $D$ since
$[E,D^{\dag}]=0$. Moreover we have $[E,H_{I}]=0$ by construction.
Our instantaneous quantum storage subspace ${\cal V}(t)$ is given
by the linear span of the following family of instantaneous
dark-states, i.e., the eigen-states of $H_{I}(t)$ with vanishing
eigenvalues
\begin{equation}
|{D}_{m,n}(t)\rangle=\frac{1}{\sqrt{m!n!}}D^{\dagger m}E^{\dagger n}|{\bf 0}%
\rangle,   \label{dark}
\end{equation}
where ${|{\bf 0}}\rangle=|0\rangle_{p}\otimes|{\bf b}\rangle\equiv
|0\rangle_{p}\otimes|b,b,...,b\rangle$ represents the ground state
of the total coupled system with each atom being in the ground
state $|b\rangle$ and the quantum probe field being in the vacuum
state $|0\rangle_{p}$. It is easy to prove that any other
dark-state polariton operator can be expressed as a linear
combination of $D$ and $E$. Notice that one can introduce the
bright-state polariton operator: $B=a\sin\theta+C\cos\theta$, which
can be used to generate eigen-states involving the excited state
$|a\rangle$. However, as shown in Ref. \cite{Sun-prl}, under the
adiabatic manipulations, these states will not get coupled to the
above constructed dark states (\ref{dark}). Of course, the states
obtained by applying $D$ are not absolutely dark since the excited
state can spontaneously decay.

To study the geometric quantum information storage based on the
above zero-eigenvalue dark states (\ref{dark}), we now consider the
quantum evolution in ${\cal V}(t)$ caused by the adiabatic change
of the parameters $\Omega_{k}(t)$. The adiabatic condition is here
given by \cite{zee,prd}
\begin{equation}
\frac{g\sqrt{N}x_{k}}{(\sqrt{g^{2}N+\Omega^{2}})^{3}}\ll1,\ \
x_{k}=\dot {|\Omega_{k}|},\Omega\delta_{k}
\end{equation}
for $k=1,2$. Let us consider a state vector
$|\Phi(t)\rangle=\sum_{m,n}c_{mn}(t)|{D}_{m,n}(t)\rangle$ belonging
to ${\cal V}(t)$. A straightforward calculation gives the matrix
equation \cite{zee,prd} for the
coefficients $c_{mn}(t)$: 
$\partial_{t}{\bf C}(t)={\bf K}(t){\bf C}(t)$, 
where the vector ${\bf C}(t)$ of coefficients and the connection matrix $%
{\bf K}(t)$ are respectively defined by ${\bf C}%
(t)=(c_{00}(t),c_{01}(t),...;c_{10}(t),c_{11}(t),...)^{T}$, and ${\bf K}%
(t)_{m,m^{\prime},n,n^{\prime}}=-\langle{D}_{m^{\prime},n^{\prime}}(t)|%
\partial_{t}{D}_{m,n}(t)\rangle\,(m,m^{\prime},n,n^{\prime}=0,1,2,\ldots)$.

The quantum storage space ${\cal V}(t)$ is, in the considered
limit, an infinite dimensional one. Thus in general it is difficult
to write down the relevant connection matrix ${\bf K}(t)$
explicitly. On the other hand the adiabatic quantum evolution in
${\cal V}(t)$ can be reduced, i.e., this space splits into
dynamically invariant finite-dimensional sectors. Let us explain
this point now.

We first observe that the following dynamical commutation relations hold $%
f_{DD}(t):=[D,\partial_{t}D^{\dag}]=-i\sin^{2}\theta(\delta_{1}\cos^{2}%
\kappa+\delta_{2}\sin^{2}\kappa)$; $\;$
$f_{ED}(t):=[E,\partial_{t}D^{\dag
}]=-\dot{\kappa}\sin\theta+i(\delta_{2}-\delta_{1})\sin\theta\cos\kappa\sin\kappa$; $\;$ $f_{DE}(t)=[-f_{ED}(t)]^{\ast}:=[D,\partial_{t}E^{\dag}]$%
; $\;$ $f_{EE}(t):=[E,\partial_{t}E^{\dag}]=-i(\delta_{1}\sin^{2}\kappa
+\delta_{2}\cos^{2}\kappa)$. Using these relations for $l^{\prime}\geq m\geq0
$ and $l\geq n\geq0$, we obtain
\begin{eqnarray}
& \langle{D}_{l^{\prime}-m,m}(t)|\partial_{t}{D}_{l-n,n}(t)\rangle  \nonumber
\\
& =\delta_{l^{\prime},l}\delta_{m,n}[(l-m)f_{DD}(t)+mf_{EE}(t)] \\
& +\delta_{l^{\prime},l}\delta_{m,n-1}\sqrt{(m+1)(l-m)}f_{DE}(t)
\nonumber
\\
& +\delta_{l^{\prime},l}\delta_{m,n+1}\sqrt{m(l-m+1)}f_{ED}(t).
\nonumber \label{conn}
\end{eqnarray}
Now it is clear from this expression that the total space ${\cal V}(t)$ can
be decomposed into a direct sum of sub-spaces: ${\cal V}(t)=\oplus
_{k=0}^{\infty}{\cal V}_{l}(t)$, where ${\cal V}_{l}(t)={\rm {span}}\{|{D}%
_{l-m,m}(t)\rangle|m=0,1,...,l\}$ has dimension ($l+1$). Notice that each $%
{\cal V}_{l}(t)$ is an invariant sub-space under the adiabatic manipulation,
i.e., if $|\Phi_{l}(0)\rangle\in{\cal V}_{l}(t)$, then $|\Phi_{l}(t)\rangle=%
\sum_{m}c_{m}^{(l)}(t)|{D}_{l-m,m}(t)\rangle \in{\cal V}_{l}(t)$. The
restricted dynamics in ${\cal V}_{l}(t)$ is governed by the reduced dynamic
equation $\partial_{t}{\bf C}_{l}(t)={\bf K}_{l}(t){\bf C}_{l}(t)$, where
the sub-coefficient vector ${\bf C}_{l}(t)$ and the reduced connection
matrix ${\bf K}_{l}(t)$ are respectively given by ${\bf C}%
_{l}(t)=(c_{0}^{(l)}(t),c_{1}^{(l)}(t),...,c_{l}^{(l)}(t))^{T}$, and ${\bf K}%
_{l}(t)=(-\langle{D}_{l-m,m}(t)|\partial_{t}{D}_{l-n,n}(t)%
\rangle)_{m,n=0,1,2,...,l}$. The solution ${\bf C}_{l}(t)={\bf W}_{l}(t)$ $%
{\bf C}_{l}(0)$ formally determines the non-abelian holonomy ${\bf W}_{l}(t)=%
{\bf T}\exp [\int{\bf K}_{l}(t)dt]$, where ${\bf T}$ is the
time-ordering operator. This non-abelian holonomy is non-diagonal
and thus can mix different instantaneous eigen-states
$|{D}_{l-m,m}(t)\rangle$ $(m=0,...,l)$ inducing in this way a truly
non-abelian gauge structure.

In the following discussion, we consider the simplified model
related to the above system as shown in Fig. 1: $\delta_{1,2}$ $
\equiv0$. Such a system has only two controllable parameters
$\Omega_{1,2}$ and can be readily realized experimentally.
Mathematically the sub-connection can be simplified as
\begin{eqnarray}
{\bf K}_{l}(t) &=& (-\langle{D}_{l-m,m}(t)|\partial_{t}{D}_{l-n,n}(t)%
\rangle)_{m,n=0,1,2,...,l} \\
&\equiv& \dot{\kappa}\sin\theta\,{\bf K}_{l}^{(0)},  \nonumber
\end{eqnarray}
where ${\bf K}_{l}^{(0)}$ is a constant matrix whose ($m,n$) entry is

\begin{equation}
\delta_{m,n-1}\sqrt{(m+1)(l-m)}-\delta_{m,n+1}\sqrt{m(l-m+1)}.
\end{equation}
In this case the time-ordering becomes irrelevant and the non-abelian
holonomy can be explicitly computed. In fact we have
\begin{equation}
{\bf W}_{l}(t)=\exp[\phi(t){\bf K}_{l}^{(0)}],
\end{equation}
where $\phi(t)=\int\dot{\kappa}\sin\theta dt$. By noticing that
${\bf K}_{l}^{(0)}$ is proportional to $(E^\dagger D-\rm{h.c.})$
restricted ${\cal V}_{l}(t),$ we find that
 this non-abelian holonomy can be rather easily cast in a diagonal form
by introducing a new instantaneous basis.{\bf \ }Let
us introduce the new set of dark-state polariton operators
\begin{equation}
D^{\prime}=\frac{1}{\sqrt{2}}(iD+E),E^{\prime}=\frac{1}{\sqrt{2}}(-iD+E),
\label{DEP}
\end{equation}
and the associated dark states $|{D}_{l-n,n}^{\prime}(t)\rangle
$=$\frac{D^{\prime\dagger l-n}E^{\prime\dagger n}}{\sqrt{
(l-n)!n!}}|{\bf 0}\rangle$. A straightforward calculation then
gives a diagonal connection matrix ${\bf
K}_{l}^{\prime}(t)=-i\dot{\kappa}\sin\theta{\rm
diag}\,(l,l-2,...,-l)$ and the corresponding  holonomy
\begin{equation}
{\bf W}_{l}^{\prime}(t)={\rm diag}\,(e^{-il\phi(t)},e^{-i(l-2)\phi
(t)},...,e^{il\phi(t)}).   \label{abelianized}
\end{equation}

Generally in EIT based quantum information storage protocols, the
Rabi frequencies $\Omega_{1,2}$ of the two classical control fields
are initially set to a very large value compared to $g\sqrt{N}$ and
then decreased independently and adiabatically (e.g., as shown in
Fig. 2). Thus $
\theta(t=0)\rightarrow0$ and $D(0)\rightarrow a$. The initial state $%
|\Phi(0)\rangle$ $=\sum_{l}c_{0}^{(l)}(0)|l\rangle_{p}\otimes|{\bf
b}\rangle$
can be written as %
$|\Phi(0)\rangle\equiv\sum_{l,m}c_{m}^{\prime(l)}(0)|D_{l-m,m}^{\prime
}(0)\rangle$, 
relative to the new basis $|{D}_{l-m,m}^{\prime}(t)\rangle$ with
the coefficients
\begin{equation}
c_{m}^{\prime(l)}(0)=\frac{(-1)^{l-m}\sqrt{l!}c_{0}^{(l)}(0)}{(i\sqrt{2})^{l}%
\sqrt{m!(l-m)!}}.
\end{equation}
and 
$|\Phi(t)\rangle=\sum_{l,m}c_{m}^{\prime(l)}(t)|{D}_{l-m,m}^{\prime}(t)%
\rangle$, 
where $c_{m}^{\prime(l)}(t)=\exp[-i(l-2m)\phi(t)]c_{m}^{\prime(l)}(0)$. When $%
\Omega_{1,2}$ become negligible compared to $g\sqrt{N}$ at time $\tau$, $%
\theta(\tau)\rightarrow\pi/2$ and $D(\tau)\rightarrow-C_{1}\cos\kappa
-C_{2}\sin\kappa$. This means that the quantum information, initially
encoded in photonic states, has been transferred and written to atomic
collective excitations. This accomplishes the quantum information storage
protocol.

\begin{figure}[h]
\hspace{30pt}\includegraphics[width=4cm,height=2.5cm]{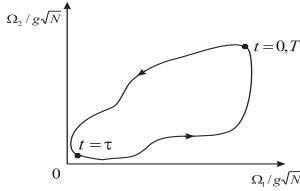}
\caption{Cyclic evolution of the parameters $\Omega _{1,2}$. At time $t=\protect\tau$, $\Omega _{1,2} \ll g\protect\sqrt{N}$; at time $%
t=0$ or $t=T$, $\Omega _{1,2} \gg g\protect\sqrt{N}$.}
\end{figure}

In order to recover the stored information one needs to drive
adiabatically the system parameters $\Omega_{1,2}$ along a cyclic
evolution such that at the time $T$ the condition $\Omega_{1,2}\gg$
$g\sqrt{N} $ is satisfied in order to guarantee $\theta(T)$
$\rightarrow0$ (see the Fig. 2). At the intermediate times
$t\in(0,\,T)$ quantum information is encoded in a combination of
photonic and atomic collective excitations. In general, if one
wants to recover exactly the initial state after that the adiabatic
loop has been completed, she has to perform a unitary
transformation to get rid of the effect of the non-abelian Berry
phase factor. In particular, for a cyclic evolution of the
parameters $\Omega_{1,2}$ if
\begin{equation}
\phi(T)=\int_{0}^{T}\dot{\kappa}\sin\theta dt=2j\pi
\end{equation}
($j$ is an integer), it then follows that $c_{m}^{\prime(l)}(T)=c_{m}^{%
\prime(l)}(0)$. In this case the system state at the final time
$T$ coincides with the initial state $|\Phi(0)\rangle.$

We are now in the position, before concluding, to make a few
comments on the relations between the results presented in this
paper and the general holonomic approach to quantum information
processing \cite{Zana,duan-science}. In that approach information
is encoded in degenerate eigenstates of a parametric family of
Hamiltonians, and in the generic case universal quantum computation
\cite{qip} can be achieved by resorting to non-abelian holonomies
only \cite{Zana}. By regarding the non-trivial holonomy one gets
after an adiabatic loop as a designed quantum state transformation,
rather than something  one wants to get rid of, it should be then
evident that the EIT-based scheme here discussed represents an
instance of such general
strategy. For example the one exciton space ${\cal V}_{1}$ can encode one {\em qubit}: $|%
{\bf 0}\rangle:=E^{\prime\dagger}|0\rangle,\,|{\bf
1}\rangle:=D^{\prime\dagger}|0\rangle. $
In this language the
transformation (\ref{abelianized}) is nothing but a diagonal
phase-shift \cite{qip}. In order to get non-diagonal single-qubit
operations one would have to relax the condition
$\delta_{1}=\delta_{2}=0.$ Encoding many-qubit states and enacting
a controllable geometric coupling between them -- as required for
realizing universal computations -- along with the robustness
of the scheme against the various sources of errors is a more complex
problem that calls for further investigations.

 {\ We acknowledge the support of
the CNSF (grant No. 90203018), the Knowledge Innovation Program
(KIP) of the Chinese Academy of Sciences and the National
Fundamental Research Program of China (No. 001GB309310). P.Z.
gratefully acknowledges financial support by Cambridge-MIT
Institute Limited and by the European Union project TOPQIP
(Contract IST-2001-39215).}

\end{document}